\documentclass{article}
\usepackage{amsmath}
\usepackage{wrapfig}
\usepackage{pgfplots}
\usepackage{enumerate}
\usepackage{amssymb}
\usepackage{algorithm}
\input{Qcircuit}
\usepackage[noend]{algpseudocode}
\usepackage{amsmath}
\usepackage[justification=centering]{caption}
\numberwithin{equation}{section}
\usepackage{authblk}
\usepackage{hyperref}

\newcommand*{\doi}[1]{\href{http://dx.doi.org/#1}{doi: #1}}

\providecommand{\keywords}[1]{\textbf{\textit{Keywords: }} #1}
\def\PACSname{\textbf{PACS}\enspace}
\def\PACS#1{\par\addvspace\medskipamount{\rightskip=0pt plus1cm
\def\and{\ifhmode\unskip\nobreak\fi\ $\cdot$
}\noindent\PACSname\ignorespaces#1\par}}
\title{A Quantum Algorithm for Testing Juntas in Boolean Functions}
\date{}

\author[1]{
  Khaled El-Wazan  \thanks{khaled\_elwazan@alex-sci.edu.eg}
}

\author[1,2]{
  Ahmed Younes \thanks{ayounes@alexu.edu.eg}
}

\author[1]{
 	 \\ S. B. Doma \thanks{sbdoma@yahoo.com}
}

\affil[1]{Department of Mathematics and Computer Science, Faculty of Science, Alexandria University, Egypt}
\affil[2]{School of Computer Science, University of Birmingham, Birmingham, B15 2TT, United Kingdom}

\begin{document}

\maketitle

\begin{abstract}
\label{abs}
Given a Boolean function $f$ provided as a black-box with $n$ variables, this paper will propose a quantum algorithm for testing if a certain variable is junta or $\epsilon$-far from being junta. The proposed algorithm constructs another black-box using two copies of the provided black-box. The constructed black-box is used with the partial diffusion operator in an amplitude amplification technique to test whether the variable being tested is junta or not, using $\mathcal{O}(\sqrt{2^n})$ queries to the constructed black-box. The proposed algorithm considers a Boolean function of general form, contrary to relevant algorithms proposed by others.



\end{abstract}


\section{Introduction}
\label{intro}
A $k$-junta function is a function of $n$ input variables which depends on at most $k$ unknown variables out of the $n$ variables for this function, where $k\leq n$. The problem of finding whether a Boolean function is a $k$-junta function or $\epsilon$-far from being a $k$-junta function is, for example, considered a typical problem in machine learning in which there is often no way to explicitly discriminate relevant features to the learning process of an unknown function from other irrelevant features \cite{blum1,blum2}. Therefore, it is necessary useful to use an adequate algorithm for testing whether an unknown function is a $k$-junta function or not, before engaging in running any $k$-junta learning algorithm.

Quantum computers \cite{feynman,deutsch} are inherently probabilistic devices that promise to execute some types of computations more powerfully than classical counterpart. Many quantum algorithms have been presented, for example, P. Shor introduced a quantum
algorithm \cite{shor} for factorizing a composite integer into its prime factors in polynomial time. L. Grover gave an algorithm \cite{grover} for searching an unstructured list of $2^n$ items with quadratic speed-up over algorithms running on classical computers.

Bernstein-Vazirani's algorithm \cite{Bernstein-Vazirani} is one of the earliest quantum algorithms, which dealt with the oracle identification problem, where it is required to identify an unknown linear Boolean function given as a black-box. The complexity of such problem is measured by how many queries are required to know the exact form of the function itself. Using classical computations, it would require $\mathcal{O}(n)$ oracle calls, however, it is solved using only a single query to the oracle using Bernstein-Vazirani's quantum algorithm.

In 2007, At\i c\i~and Servedio introduced a quantum $\delta$-property tester \cite{atici-servedo} for $k$-junta Boolean functions using $\mathcal{O}(k/\delta)$ quantum queries and based on Fourier sampling \cite{Bernstein-Vazirani}. As well, they introduced an algorithm for learning a $k$-junta to accuracy $\delta$ that uses $\mathcal{O}(\delta^{-1}k \log{k})$ quantum examples and $\mathcal{O}(2^k \log{\delta^{-1}})$ random examples.


Floess \textit{et al.} introduced a Bernstein-Vazirani based algorithm \cite{floess} for finding the set of input variables that a Boolean function depends on, followed by an amplitude amplification algorithm \cite{grover} to increase the success probability of finding those variables. Floess \textit{et al.}'s junta tester runs in $\mathcal{O}(2^n)$ oracle calls. As well, Floess \textit{et al.} proposed quantum algorithms for learning which variable resided in a linear, quadratic or cubic terms in the function,  only with the assumption that each variable appears in, at most, one term.

Li and Yang presented a quantum algorithm \cite{yang} which evaluates the influence of a variable on a Boolean function, using $\mathcal{O}(1)$ steps of Bernstein-Vazirani algorithm. Li and Yang also discussed a probabilistic algorithm for learning quadratic and cubic functions of simple forms.

Ambainis \textit{el al.} introduced a quantum algorithm \cite{Ambainis} for testing $k$-junta Boolean functions. Ambainis \textit{el al.} quantum algorithm is based on an algorithm that solves the group testing problem \cite{sterrett}. This $k$-junta algorithm is found to require $\mathcal{\tilde{O}}(\sqrt{k/\delta })  $ oracle call to the tested function which represents a quadratic speed-up over the quantum junta property tester in \cite{atici-servedo}.

The aim of this paper is to propose an algorithm to test whether a variable $x_i$ in a Boolean function $f$ provided as a black-box with $n$ input variables is a junta variable or $\epsilon$-far from being junta. The proposed algorithm can identify whether a variable $x_i$ in the function $f$ is relevant or not, using a new function $g$ which is constructed from the given black-box. The new constructed function $g$ is, then, used with an amplitude amplification algorithm based on partial diffusion operator to increase the success probability. As well, the algorithm works on any class of Boolean functions with probability of success at least $2/3$.

The paper is organized as follows: Section~\ref{quantum-search-algorithm} depicts a quantum search algorithm with more reliable behavior for both known and unknown number of matches. Section~\ref{oracle-construction} introduces the construction of a new function $g$ using the original black-box which will facilitate the junta property testing. Section~\ref{proposed-algorithm} presents the proposed algorithm. Section~\ref{analysis} describes the performance of the proposed algorithm when testing any Boolean function regardless of its form. Section~\ref{comparison-with-floess} compares the proposed algorithm with other relevant algorithms, followed by a  general conclusion in Section~\ref{conclusion}.

\section{Quantum Search Algorithm}
\label{quantum-search-algorithm}

Let's consider having a list $L$ of $N=2^n$ items, that has an oracle $U_f$ which is used to access those items. Each item $l\in L$ is labeled with an integer $\lbrace 0, 1, ..., N-1\rbrace$ and mapped to either $0$ or $1$ according to any certain property satisfied by $l$, \textit{i.e.} $f:L\rightarrow \lbrace 0,1 \rbrace$. The search problem is to find $l\in L$ such that $f(l)=1$.

L. Grover introduced in 1996 a novel approach for solving this typical problem with quadratic speed-up over classical algorithms \cite{grover}. The algorithm proposed by Grover  exploits quantum parallelism by preparing a uniform superposition which represents all the possible $N$ items, marks the solution using phase shift of $-1$ using the oracle $U_f$ then amplifies the amplitude of the solution using inversion about the mean (diffusion operator). Grover's algorithm has shown to be optimal \cite{grover-optimal} with high success probability if there is exactly one item $l$ in the list $L$ that satisfies the oracle $U_f$, and required approximately $\pi/4\sqrt{N}$ iterations for that particular case \cite{grover}. Grover's search algorithm was generalized by Boyer \textit{et al.} for known multiple matches $M$ that satisfied the oracle $U_f$, \textit{i.e.} $\forall j,$ for which $ 1\leq j\leq M\leq 3N/4$, $f(l_j)=1$, and the generalized Grover algorithm is found to require a number of $\pi/4\sqrt{N/M}$  iterations \cite{boyer}. In addition, in the case of unknown number of matches $M$, an algorithm is presented to  find a match \cite{boyer}. It was found that the generalized Grover algorithm fails in the case of $M>3N/4$ \cite{boyer,grover-younes}.

Younes \textit{et al.} introduced a more reliable algorithm \cite{younes-miller} in the case of multiple matches than the generalized Grover algorithm, for $1\leq M\leq N$, and for fewer matches, the algorithm runs in quadratic speed-up similar to the generalized Grover algorithm. 

In the following section, Younes \textit{et al.}'s algorithm for both known and unknown number of matches $M$ will be reviewed since they will be used in our proposed algorithm.

\subsection{In Case of Known Number of Matches $M$}

\begin{figure}[H]
\begin{align*}
 \Qcircuit @C=1em @R=.7em {
  \lstick{\ket{0}} & /^n \qw & \gate{H^{\otimes n}} & \multigate{1}{U_f} & \qw & \multigate{1}{Y} & /^n \qw 	& \meter &  \qw  \\
  \lstick{\ket{0}} & \qw     & \qw   & \ghost{U_f}   & \qw & \ghost{Y}     & \qw  \gategroup{1}{4}{2}{6}{.7em}{_\}} \\
  \\
  & & & & & \lstick{O(\sqrt{N/M})} & &
 }
\end{align*}
\caption{Quantum circuit for the quantum search algorithm \cite{younes-miller}.\label{quantum-search-circuit}}
\end{figure}
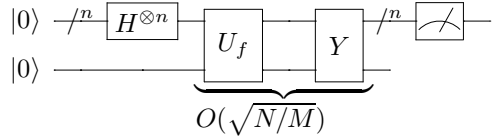

For a list $L$ of size $N=2^n$, the steps are as follows:
\begin{enumerate}

\item Prepare a quantum register with $n+1$ qubits in a uniform superposition

\begin{equation}
|\varphi\rangle = \frac{1}{\sqrt{N}}\sum_{l=0}^{N-1}\ket{l} \otimes \ket{0}.
\end{equation}

\item Iterate the algorithm for ${\pi}/{(2\sqrt{2})}\sqrt{{N}/{M}}$ times by applying the partial diffusion operator $Y$ on the state $U_f|\varphi\rangle$ in each iteration, such that it performs the inversion about the mean on a subspace of the system, where

\begin{equation}
Y=(H^{\otimes n}\otimes I)(2|0\rangle\langle 0|-I_{n+1})(H^{\otimes n}\otimes I).
\end{equation}

At any iteration $q\geq 2$, the system can be described as follows \cite{younes-miller}:
\begin{equation}
\ket{\varphi^q}=a_q\sum_{l=0}^{{N-1}}\strut^{\prime\prime}\big(|l\rangle\otimes|0\rangle\big)+b_q\sum_{l=0}^{{N-1}}\strut^{\prime}\big(|l\rangle\otimes|0\rangle\big)+c_q\sum_{l=0}^{{N-1}}\strut^{\prime}\big(|l\rangle\otimes|1\rangle\big).
\end{equation} 
where,
\begin{align}
a_q&=2\langle\alpha_q\rangle-\alpha_{q-1}, \quad
b_q=2\langle\alpha_q\rangle-c_{q-1}, \quad
c_q=-b_{q-1},
\end{align}
\begin{equation}
\langle\alpha_q\rangle=\Big(\big(1-\frac{M}{N}\big)\alpha_{q-1}+\big(\frac{M}{N}\big)c_{q-1}\Big).
\end{equation}
and $\sum\strut^\prime$ indicates the sum over all desired states, and $\sum\strut^{\prime\prime}$ indicates the sum over the undesired states.

\end{enumerate}

For this algorithm, the success probability is as follows \cite{younes-miller}:

\begin{equation}
P_s=\big(1-\cos\big(\theta\big)\big)\Big(\frac{\sin^2\big(\big(q+1\big)\theta\big)}{\sin^2\big(\theta\big)}+\frac{\sin^2\big(q\theta\big)}{\sin^2\big(\theta\big)}\Big),
\end{equation}
where $\cos\big(\theta\big)=1-M/N$; $0< \theta\leq\pi/2$, and the required number of iterations $q$ is given by:

\begin{equation}
q=\Bigl\lfloor\frac{\pi}{2\theta}\Bigr\rfloor \leq\frac{\pi}{2\sqrt{2}}\sqrt{\frac{N}{M}},
\end{equation}
where $\lfloor~\rfloor $ is the floor operation.

The algorithm of Younes \textit{et al.} \cite{younes-miller} is noted to be slower than the generalized Grover algorithm \cite{boyer} for small $M/N$ by $\sqrt{2}$, yet Younes \textit{et al.} algorithm is more reliable with high probability when handling the range $1\leq M\leq N$ than the generalized Grover algorithm that covers the range $1\leq M\leq 3N/4$.

\subsection{In Case of Unknown Number of Matches $M$}

The previous section described  Younes \textit{et al.}'s algorithm in case of known number of matches $M$ \cite{younes-miller}. However, it is difficult to apply that quantum search algorithm (even generalized Grover algorithm \cite{boyer}) without knowing the number of solutions $M$, because the algorithm is sensitive to the number of iterations which depends on the number of solutions $M$ to the problem itself. 
Younes \textit{et al.} described an algorithm \cite{younes-miller} similar to \cite{boyer}, that handles the searching problem in case of unknown number of matches $M$, which is as follows for $1\leq M \leq N$ :
\begin{enumerate}
\item Start with $m=1$ and put $\lambda=8/7$.
\item Choose a positive integer $s$ uniformly at random such that $s<m$.
\item Apply $s$ iterations of Younes \textit{et al.}'s algorithm on the state:
\begin{equation}
|\varphi \rangle = {\frac{1}{\sqrt{N}}}  \sum_{l}^{N} |l\rangle. \nonumber
\end{equation}
\item Measure the register assuming its output is $t$.
\item If f(t)=1, then the problem is solved and exit. Otherwise, set $m$ to the minimum between $\lambda m$ and $\sqrt{N}$ and go back to Step $2$.
\end{enumerate}

Grover algorithm is employed in  \cite{boyer} to find solutions when $M$ is unknown, and the estimated solutions grow exponentially when $M> 3N/4$ \cite{boyer, grover-younes}.
However, employing Younes \textit{et al.}'s algorithm for finding a solution when the number of solutions is $M$, requires $\mathcal{O}(\sqrt{N/M})$ when $1\leq M\leq N$ which is better compared to using the generalized Grover algorithm \cite{boyer}.

\section{Constructing the Oracle $U_g$}
\label{oracle-construction}

In this section, the oracle $U_f$ will be used in a specific way to fit the purpose of finding whether the variable $x_i$ in the function $f$ is junta or not before applying the amplification algorithm. 
In addition, We will provide a method to view and analyze the modification proposed.

\subsection{Variable Negation}

Any Boolean function $f$ in positive polarity Reed-Muller form  \cite{reed-muller} of $n$ input variables, and $N=2^n$ can be written as follows:

\begin{align}
f(x_0,x_1,\cdots,x_i,\cdots,x_{n-1})=\bigoplus_{q=0}^{N-1}b_q P_q \textit{ , }
\label{generalformfun}
\end{align}
where
\begin{align*}
P_q&: \emph{product term} \\
b_q&=
\begin{cases}
0:\emph{ product term $P_q$ does not exist}\\
1: \emph{product term $P_q$ exists}
\end{cases}.
\end{align*}

Let's define $f_{\bar{x}_i}$ such that,
\begin{equation}
f_{\bar{x}_i}=f(x_0,x_1,\cdots,x_{i-1},\bar{x}_i,x_{i+1},\cdots,x_{n-1}) \textit{ , }
\end{equation}
and $g_{ff_{\bar{x}_{i}}}$ as follows,

\begin{align}
g_{ff_{\bar{x}_i}}&=f\oplus f_{\bar{x}_i} \nonumber \\
&=g(x_0,x_1,\cdots,x_{i-1},x_{i+1},\cdots,x_{n-1})  \nonumber \\
&=\bigoplus_{\beta=2^{n-i-1}}^{Q-1}c_\beta P_\beta,
\label{gff}
\end{align}
where $Q=2^{n-1}$, $c_\beta=b_\mu$ and $\mu$ is the bit representation of the $\beta$th term of size $n$ bits but only with the bit at position $i$ equals to $1$, \textit{i.e.} $\mu=\beta_0 \beta_1 \beta_2 \cdots \beta_{i-1} 1 \beta_{i+1}\cdots \\ \beta_{n-1}$. 
It should be noted that $g_{ff_{\bar{x}_i}}$  will decompose the function $f$ to a lower order general function and the variable $x_i$ in question will disappear from the definition of $g_{ff_{\bar{x}_i}}$.

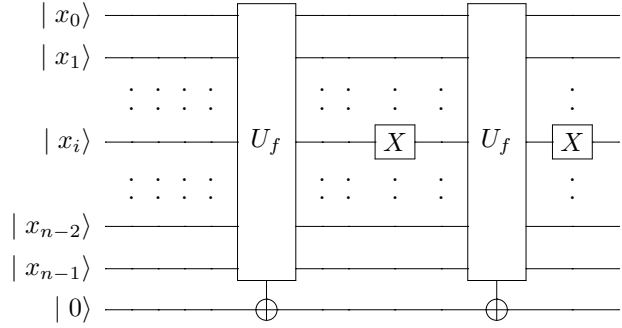
\begin{figure}[H]
\begin{align*}
\Qcircuit @C=1em @R=.7em  {
\lstick{\mid x_0 \rangle}& \qw & \qw & \qw & \qw & \multigate{8}{U_f} & \qw  & \qw & \qw & \qw & \multigate{8}{U_f} & \qw & \qw  \\
\lstick{\mid x_1 \rangle}& \qw & \qw & \qw & \qw & \ghost{U_f}& \qw  & \qw & \qw & \qw & \ghost{U_f} & \qw & \qw\\
& . & . & . & . &  & .  & . & . & . &   & . &   \\
& . & . & . & . &  & .  & . & . & . &   & . &   \\
\lstick{\mid x_i \rangle} & \qw & \qw & \qw & \qw & \ghost{U_f}& \qw  & \qw & \gate{X} & \qw & \ghost{U_f} & \gate{X} & \qw\\
& . & . & . & . &  & .  & . & . & . &   & . &   \\
& . & . & . & . &  & .  & . & . & . &   & . &   \\
\lstick{\mid x_{n-2} \rangle}& \qw & \qw & \qw & \qw & \ghost{U_f}& \qw  & \qw & \qw & \qw & \ghost{U_f} & \qw & \qw\\
\lstick{\mid x_{n-1} \rangle}& \qw & \qw & \qw & \qw & \ghost{U_f}& \qw  & \qw & \qw & \qw & \ghost{U_f} & \qw & \qw\\
\lstick{\mid 0 \rangle}& \qw & \qw & \qw & \qw &  \targ{-1} \qwx & \qw  & \qw & \qw & \qw & \targ \qwx & \qw & \qw \\
}
\end{align*}
\caption{ A quantum circuit for the proposed oracle $U_g$.\label{fig:proposed-oracle-circuit}}
\end{figure}

A quantum circuit for the oracle $U_g$ can be constructed as follows: $U_g=U_{f_{\bar{x}_i}}U_f$, if the variable exists in at least one term in the function $f$, and if $x_i$ is junta, then $U_{f_{\bar{x}_i}}=U_f$ and then $U_g=I_n$, where $I_n$ is the identity matrix of size $2^n\times 2^n$. An illustration  of this circuit is shown in Figure \ref{fig:proposed-oracle-circuit}.

It is noted that the variable $x_i$ in question will disappear from the general expression of $g_{ff_{\bar{x}_i}}$. For further elaboration, let's study the case of a general function of 2 inputs which represents all possible 2-variable functions $f(x_0,x_1)$ in positive polarity Reed-Muller form:

\begin{align}
f(x_0,x_1)=b_0\oplus b_1x_1\oplus b_2x_0\oplus b_3x_0x_1.
\label{function_order2}
\end{align}

Let's assume that the variable in question is $x_1$, then

\begin{align}
f(x_0,\bar{x}_1)=b_0\oplus b_1\bar{x}_1\oplus b_2x_0\oplus b_3x_0\bar{x}_1.
\end{align}

It is known that $\bar{x}_i=x_i\oplus 1$, so that

\begin{align}
f(x_0,\bar{x}_1)&=b_0\oplus b_1  (1\oplus x_1) \oplus b_2x_0\oplus b_3 x_0 (1\oplus x_1) \nonumber \\  &=
b_0\oplus b_1 x_1\oplus b_1\oplus b_2 x_0\oplus b_3 x_0 x_1 \oplus b_3 x_0.
\end{align}

Let's define a new function $g_{ff_{\bar {x}_i}}$, such that $g_{ff_{\bar x_1}}=f(x_0,x_1) \oplus f(x_0,\bar{x}_1)$, as follows:

\begin{align}
g_{ff_{\bar x_1}  }&=f(x_0,x_1)\oplus f(x_0,\bar{x}_1) \nonumber \\
&=b_0\oplus b_1x_1\oplus b_2x_0\oplus b_3x_0x_1\oplus b_0\oplus b_1 x_1\oplus b_1\oplus b_2 x_0\oplus b_3 x_0 x_1 \oplus b_3 x_0 \nonumber  \\
&=b_1\oplus b_3 x_0.
\end{align}

Let's study the case of a simple general function of $3$ inputs, which represents all possible 3-variable functions, $f(x_0,x_1,x_2)$ in Reed-Muller form:

\begin{align}
f(x_0,x_1,x_2)&=b_0\oplus b_1 x_2\oplus b_2x_1\oplus  b_3x_1x_2\oplus b_4x_0\oplus b_5x_0x_2 \nonumber \\ & \oplus b_6x_0x_1\oplus b_7x_0x_1x_2
\label{function_order3}.
\end{align}

Let's assume that the variable in question is $x_0$, the function $f(\bar{x}_0,x_1,x_2)$ will be as follows:

\begin{align}
f(\bar{x}_0,x_1,x_2)=b_0\oplus b_1x_2\oplus b_2x_1\oplus x_1x_2\oplus b_4x_0\oplus b_4\oplus b_5x_0x_2 \oplus \nonumber  \\ b_5x_2\oplus b_6x_0x_1 \oplus b_6x_1 \oplus b_7x_0x_1x_2\oplus b_7x_1x_2.
\end{align}

Defining $g_{ff_{\bar{x}_i}}$, such that $g_{ff_{\bar{x}_0}} = f(x_0,x_1,x_2) \oplus f(\bar{x}_0,x_1,x_2)$, will yield

\begin{equation}
g_{ff_{\bar{x}_0}} = b_4\oplus b_5x_2\oplus b_6x_1\oplus b_7x_1x_2.
\label{order3}
\end{equation}

\section{The Proposed Algorithm}
\label{proposed-algorithm}

Any general Boolean function can be defined in terms of the variable $x_i$ as follows:
\begin{equation}
f(x_0,x_1,...,x_{n-1})=f_{+x_i}\oplus f_{-x_i},
\label{anotherGeneral}
\end{equation}
where $f_{+x_i}$ are the terms in the function $f$ which contain the variable $x_i$, and $f_{-x_i}$ are the terms in the function $f$ which do not contain that variable $x_i$.

When preparing the function $g=f\oplus f_{\bar{x}_i} $, using the new general definition 
\begin{align}
g&=f\oplus f_{\bar{x}_i} \nonumber \\
&=(f_{+x_i}\oplus f_{-x_i} )\oplus (f_{+\bar{x}_i} \oplus f_{-x_i}) \nonumber \\
&=f_{+x_i} \oplus f_{+\bar{x}_i},
\label{anotherGeneralReduction}
\end{align}
where $f_{+\bar{x}_i}$ are the terms in the function $f$ which had $x_i$ and are decomposed to lower order terms not containing $x_i$. 
Then the problem is converted to finding whether $g$ has at least one solution $(g\neq 0)$ or not.

In this section, we will propose the algorithm to test whether a variable $x_i$ in the Boolean function $f$ is a junta variable or $\epsilon$-far from being a junta variable, utilizing Younes \textit{et al.}'s algorithm for unknown number of matches and the property of the $U_{g}$ oracle discussed in Section \ref{oracle-construction}. 
The steps of the proposed algorithm is as follows:

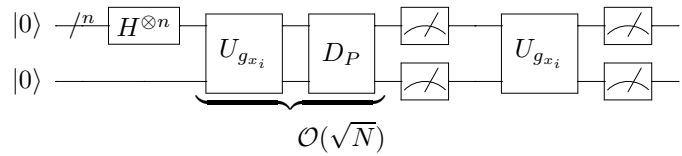
\begin{figure}[H]
\begin{align*}
 \Qcircuit @C=1em @R=.7em {
  \lstick{\ket{0}} & /^n \qw & \gate{H^{\otimes n}} & \multigate{1}{U_{g_{x_i}}} & \multigate{1}{D_P}	&   \meter & \qw & \multigate{1}{U_{g_{x_i}}} & \meter  & \qw \\
  \lstick{\ket{0}}  & \qw    & \qw   &   \ghost{U_{g_{x_i}}}  & \ghost{D_P}   & \meter & \qw & \ghost{U_{g_{x_i}}} & \meter & \qw \gategroup{1}{4}{2}{5}{.7em}{_\}}\\
  \\
  & & & &\mathcal{O}(\sqrt{N}) &
 }
\end{align*}
\label{proposed-algorithm-circuit}
\caption{Quantum circuit for the proposed algorithm.}
\end{figure}

\begin{enumerate}

%

\item Prepare $g_{x_i}=f\oplus f_{\bar{x}_i}$.
\item Test whether $g_{x_i}$ has a constant term as follows:
\begin{enumerate}
\item Prepare a vector $v=(0,0,0,...,0)$ of size $n$ such that all qubits are equal to $\vert0\rangle$.
\item If $g_{x_i}(v)=1$, which means that the vector $v$ is a solution for the function $g_{x_i}$ and this implies that $x_i$ exists in a linear term in the function $f$. The variable $x_i$ will be flagged not junta and then exit.
\end{enumerate}

%

\item Start with $m=1$ and $\lambda=8/7$, where $\lambda$, such that $1\leq \lambda \leq 4/3$.
\item \label{step} Choose a positive integer $s$ uniformly at random such that $0\leq s \leq m-1$.

\item Iterate  $s$ iterations of Younes \textit{et al.}'s algorithm on the state:
 \begin{equation}
  |\varphi \rangle = {\frac{1}{\sqrt{N}}}  \sum_{l=0}^{N-1} |l\rangle \nonumber.
 \end{equation}
 
\item Measure the register assuming its output is $t_n$.
\item If $g_{x_i}(t_n)=1$, then $x_i$ is considered to be not junta and exits.
\item Set $m$ to the minimum between $\lambda m$ and $\sqrt{N}$ and go back to Step ~\ref{step}.

\end{enumerate}

\section{Analysis of the Proposed Algorithm}
\label{analysis}

In this section, we will discuss the behavior of the proposed algorithm with the suggested oracle modification mentioned in Section~\ref{oracle-construction}, assuming that the oracle $U_f$ representing the function $f$ is a black-box oracle.

\subsection{Affine Functions}
An affine Boolean function $f$ with $n$ inputs can be generally represented as follows:
\begin{align}
f(x_0,x_1,x_2,\cdots,x_{n-1}) &= c_0x_0\oplus c_1x_1\oplus \cdots \oplus c_{n-1}x_{n-1} \oplus c_n \nonumber \\
&=\bigoplus_{i=0}^{n-1}c_ix_i\oplus c_n,
\end{align}
where the coefficient $c_i$ decides whether the term that has the variable $x_i$ exists in the definition of the function $f$ or not, \textit{i.e.} 

\begin{equation}
c_i=
\begin{cases}
      0, \emph{ if $x_i$ is junta}  \\
      1, \emph{otherwise}
\end{cases},
\label{earlier}
\end{equation}
and $c_n$ describes generally the affinity of the Boolean function, \textit{i.e.}

\begin{equation}
c_n=
\begin{cases}
      0, \emph{ if the function f is not affine}  \\
      1, \emph{ if the function f is affine}
\end{cases}.
\end{equation}


A linear Boolean function is defined as follows:
\begin{align}
f(x_0,x_1,\cdots,x_{n-1}) &= c_0x_0\oplus c_1x_1\oplus \cdots \oplus c_{n-1}x_{n-1}  \nonumber \\
&=\bigoplus_{i=0}^{n-1}c_ix_i,
\end{align}
where $c_i$ is as described in Equation ~\eqref{earlier} and $c_n=0$.



Suppose that the black-box oracle represents an affine Boolean function. This will guarantee that $g_{x_i}$ will be a constant function with one of the following:

\begin{enumerate}

\item  If the variable $x_i$ is not junta, \textit{i.e.} $c_i=1$, the resultant function will be as follows:
\begin{align}
g_{x_i}&=1,
\end{align}

which is a constant function that could be easily identified using one evaluation of the function $g_{x_i}$ as described in the proposed algorithm, \textit{i.e.} $O(1)$. 

\item If the variable $x_i$ is junta, \textit{i.e.} $c_i=0$, the resultant function will be as follows:
\begin{equation}
g_{x_i}=0,
\label{constant}
\end{equation}

which is a constant function with no solutions which requires $\mathcal{O}(\sqrt{N})$ oracle calls using the proposed algorithm.

\end{enumerate}


\subsection{Nonlinear Functions}

Suppose the algorithm is operating on a general nonlinear Boolean function $f$ which can represented as follows:

\begin{equation}
f(x_0,x_1,\cdots,x_{n-1})=\bigoplus_{q=0}^{N-1}b_qP_q,
\end{equation}
such that $P_q$ is a product term composed of $\hbar_q$ variables from the set $\{x_0,x_1,\cdots, x_{n-1}\}$, such that $1<\hbar_q\leq n$, and $b_q$ dictates whether the term $P_q$ exists or not in the function definition, then we have the following:

\begin{enumerate}
\item If the variable $x_i$ being tested is a junta variable, the resultant function will be a constant function as in Equation \eqref{constant} and will require $\mathcal{O}(\sqrt{N})$ oracle calls.
\item If the variable $x_i$ is not a junta variable, it is guaranteed that the function $g_{x_i}$ will have at least one solution which will be amplified and thus will require $\mathcal{O}(\sqrt{N})$ oracle calls to be found.
\end{enumerate}

\section{Comparison with Relevant Work}
\label{comparison-with-floess}
In 2010, Floess \textit{et al.} introduced an algorithm \cite{floess} for finding the input variables which a given function being tested depend on, based on  Bernstein-Vazirani algorithm. The success probability of finding the junta variables for the given Boolean function has been further amplified using amplitude amplification algorithm. 

\paragraph{Single term Boolean function of order $m$ amplification:}
A drawback of Floess \textit{et al.} algorithm appears when the function being tested is a single product term Boolean function of $m$ variables such that $m\leq n$,  which should be an unknown fact about the function being tested, as follows:
\begin{equation}
f(x_{0},x_{1},\cdots,x_{n-1}) = \chi,
\end{equation}
where $\chi$ is a product term of $m$ variables from the set $\{x_0,x_1,\cdots, x_{n-1}\}$.  In such case, the number of iterations required for amplitude amplification algorithm must be estimated because the amplitude amplification algorithm is sensitive to the number of iterations \cite{boyer,grover-younes}. Floess \textit{et al.} showed that for a product of $m$ input variables, the required number of iterations for amplitude amplification algorithm can be estimated using a circuit which requires $\mathcal{O}(2^{m})$ oracle calls \cite{floess,floess-thesis}, which does not provide any speed-up compared to a classical algorithm counterpart. The proposed algorithm, however, introduces a quadratic speed-up compared to Floess \textit{et al.} algorithm, and will require $\mathcal{O}(\sqrt{2^m})$ oracle calls.

\paragraph{Multiple terms Boolean function:}
Whether the decomposed function $g$ is of a single term or several terms,
 the expected function calls of the proposed algorithm is $\mathcal{O}(\sqrt{N/M})$, when $M$ is $1\leq M\leq N$. However, Floess \textit{et al.} did not cover the case when the unknown function $f$ is composed of multiple terms with different degrees.

\section{Conclusion}
\label{conclusion}

This paper proposed a quantum algorithm to test if a certain variable of a given Boolean function $f$ with $n$ variables is junta or $\epsilon$-far from being junta. The Boolean function is assumed to be provided as a black-box.  It was shown that the black-box Boolean function can be used to construct another black-box with certain properties, using two copies of the given black-box. Also, it was shown that the constructed black-box will have no solutions if the variable being tested is junta, and it will have at least one solution if the variable being tested is not junta. The number of solutions of the constructed black-box is assumed to be unknown where an amplitude amplification technique that marks the solutions with entanglement is used, then a partial diffusion operator is  used to find whether the constructed black-box has at least one solution or does not have any solutions, using $\mathcal{O}(\sqrt{2^n})$ queries to the constructed black-box.

It was shown that the proposed algorithm can handle any Boolean function provided as a black-box without any restrictions on the form of the Boolean function and with success probability at least $2/3$, where the relevant work proposed by others \cite{floess} tests certain classes of Boolean functions and the success probability depends on the form of the Boolean function being tested.


\addcontentsline{toc}{chapter}{Bibliography}

\bibliography{references}
\bibliographystyle{IEEEtran}

\end{document}